\documentclass[a4paper,fleqn,usenatbib]{mnras}

\usepackage[T1]{fontenc}
\usepackage{ae,aecompl}

\usepackage{graphicx}	
\usepackage{amsmath}	
\usepackage{amssymb}	
\usepackage{subfigure}
\usepackage{multirow}
\usepackage{xcolor}

\newcommand{\gcc}{\,{\rm g \, cm}^{-3}}

\newcommand{\kel}{\, {\rm K}}

\newcommand{\msun}{\, {\rm M}_\odot}

\newcommand{\pc}{\, {\rm pc}}

\newcommand{\myr}{\, {\rm Myr}}

\newcommand{\kms}{\, {\rm km \, s^{-1}}}
\newcommand{\av}{A_{_{\rm V}}}

\newcommand{\vcol}{v_{\rm col}}

\title[Cloud-cloud collisions]{Molecular line signatures of cloud-cloud collisions}

\author[Priestley \& Whitworth]{
F. D. Priestley\thanks{Email: priestleyf@cardiff.ac.uk} and A. P. Whitworth
\\
School of Physics and Astronomy, Cardiff University, Queen's Buildings, The Parade, Cardiff CF24 3AA, UK \\
}

\date{Accepted XXX. Received YYY; in original form ZZZ}

\pubyear{2021}

\begin{document}
\label{firstpage}
\pagerange{\pageref{firstpage}--\pageref{lastpage}}
\maketitle

\begin{abstract}
Collisions between interstellar gas clouds are potentially an important mechanism for triggering star formation. {This is} because they are able to rapidly {generate} large masses of dense gas. Observationally, {cloud} collisions are often identified in position-velocity (PV) space through bridging features between intensity peaks, usually of CO emission. Using a combination of hydrodynamical simulations, time-dependent chemistry, and radiative transfer, we produce synthetic molecular line observations of {overlapping clouds that are genuinely colliding, and overlapping clouds that are just chance superpositions.} {Molecules tracing denser material than CO, such as NH$_3$ and HCN, reach peak intensity ratios of $0.5$ and $0.2$ with respect to CO in the `bridging feature' region of PV space for {genuinely} colliding {clouds}. For {overlapping} clouds {that are just chance superpositions,} the peak {NH$_3$ and HCN intensities are co-located with the CO intensity peaks.} This represents a way of confirming cloud collisions observationally, {and distinguishing them from chance alignments} of unrelated material.}
\end{abstract}
\begin{keywords}
astrochemistry -- stars: formation -- ISM: molecules -- ISM: clouds -- ISM: structure
\end{keywords}

\begin{figure*}
  \centering
  \subfigure{\includegraphics[width=\columnwidth]{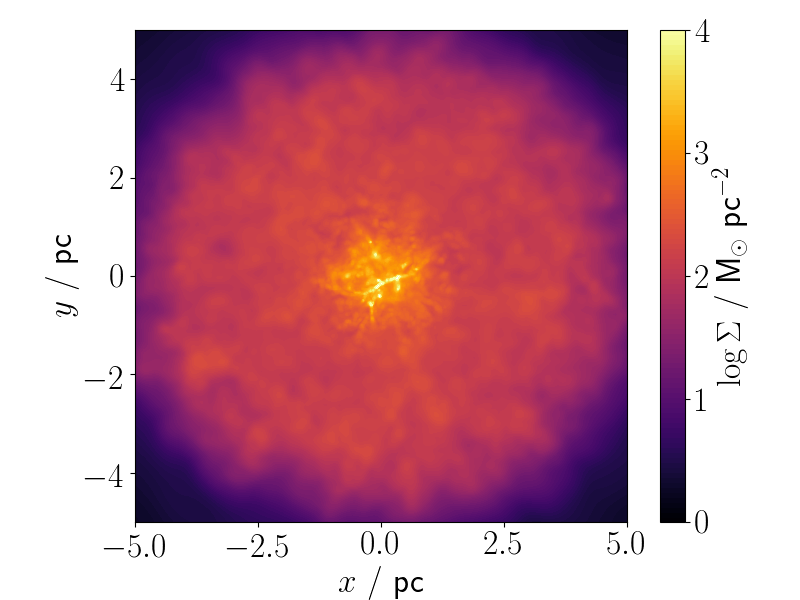}}\quad
  \subfigure{\includegraphics[width=\columnwidth]{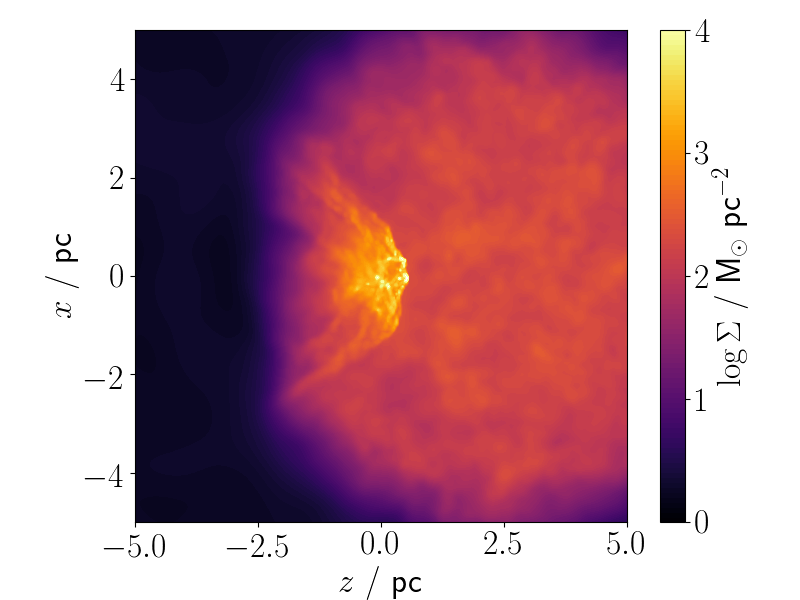}}\\
  \caption{Surface density maps {from the cloud-collision simulation at $t_{\rm end}=0.49\myr$. {\it Left panel:} looking along the collision axis, i.e. projected on the $z\!=\!0$ plane. {\it Right panel:} looking perpendicular to the collision axis, i.e. projected on the $y\!=\!0$ plane.}}
  \label{fig:dens}
\end{figure*}

\section{Introduction}

{Collisions between molecular clouds {may trigger or accelerate} star formation, {and galactic-scale} simulations suggest that cloud collisions are relatively common events \citep{tasker2009,dobbs2015}. Whilst the majority of these collisions {may} have little effect on star formation {and simply act to restructure the clouds involved,} if some fraction of collisions trigger gravitational collapse, then the observed link between global galaxy properties and local star formation \citep{kennicutt1998} can be reproduced \citep{tan2000,fujimoto2014}.

{Cloud} collisions with relative velocities $\sim 10 \kms$ appear to be fast enough to result in significant compression at the interface between {the clouds, thereby promoting} star formation, without being so fast that the clouds are entirely disrupted \citep{takahira2014,balfour2015,balfour2017,liow2020}. There are several known cases of active star-forming regions associated with multiple molecular gas components separated by velocities of this order \citep[e.g.][]{fukui2015,dobashi2019}, suggesting {that there has been collisional triggering.}}

{{Whilst} the presence of multiple velocity components in molecular line emission may be suggestive of an ongoing cloud collision, it is {not} conclusive {proof}, and additional signatures such as cloud-scale emission from shock tracers \citep{jimenez2010,cosentino2018,cosentino2020} are necessary to distinguish collisions from chance line-of-sight alignments. {The} most commonly-used {signature} is the presence of `bridging features' in position-velocity (PV) maps of CO isotopologues \citep[][and references therein]{fukui2020}. First identified in combined hydrodynamical-radiative transfer modelling of cloud collisions by \citet{haworth2015}, {bridging features} result from the deceleration of material at the interface between {two colliding clouds. This produces a shock-compressed layer which emits at velocities intermediate between the two intensity peaks corresponding to the undecelerated} clouds.

{Whilst} easy to search for, {bridging} features are only present for a {small} fraction of the cloud lifetime \citep{haworth2015b}, and become increasingly indistinct as the viewing angle approaches $90^\circ$ to the collision axis \citep{haworth2015,bisbas2017}. Additionally, the high abundance and low critical density of CO result in relatively strong emission even in low-density gas. It would {therefore not be surprising to find some emission between two intensity peaks} in PV space, {even if the peaks did not represent colliding clouds.} Whilst some proposed bridging features are remarkably similar to those in the \citet{haworth2015} simulations \citep{torii2011}, others are much less clear-cut (e.g. \citealt{issac2020}, although we note that those authors present a significant quantity of other evidence in favour of a cloud collision).}

{Previous theoretical work has focused on the CO($J\!\!=$1-0) rotational transition, specifically the $^{12}$CO isotopologue \citep{haworth2015}, and {on} atomic fine structure lines \citep{bisbas2017}, which predominantly trace lower-density gas. At higher densities, such as those found in the interface between colliding clouds, optical depth effects and depletion onto grain surfaces result in a poor correlation between CO intensity and gas mass \citep[e.g.][]{priestley2020}. {Moreover} the low abundance of atomic species {in dense gas} makes {them} unsuitable as tracers {of shock-compressed layers}. In this paper, we use combined hydrodynamical, chemical and radiative transfer models to investigate line emission from molecules {requiring higher excitation densities as potential signatures of} ongoing collisions between molecular clouds.}

\begin{figure*}
  \centering
  \subfigure{\includegraphics[width=0.3\textwidth]{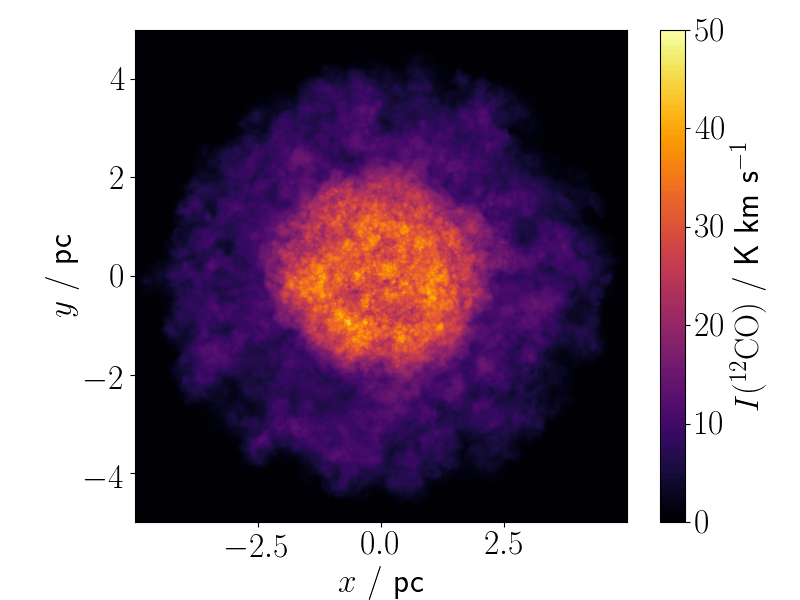}}\quad
  \subfigure{\includegraphics[width=0.3\textwidth]{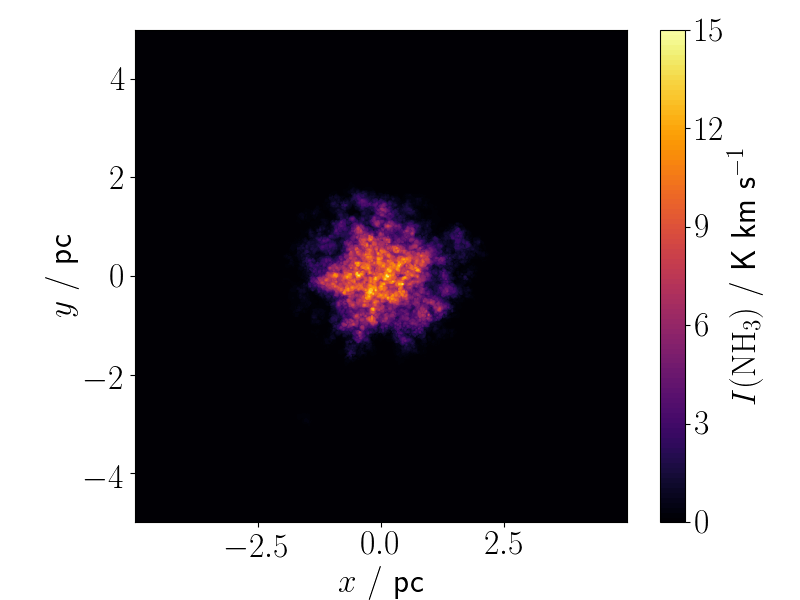}}\quad
  \subfigure{\includegraphics[width=0.3\textwidth]{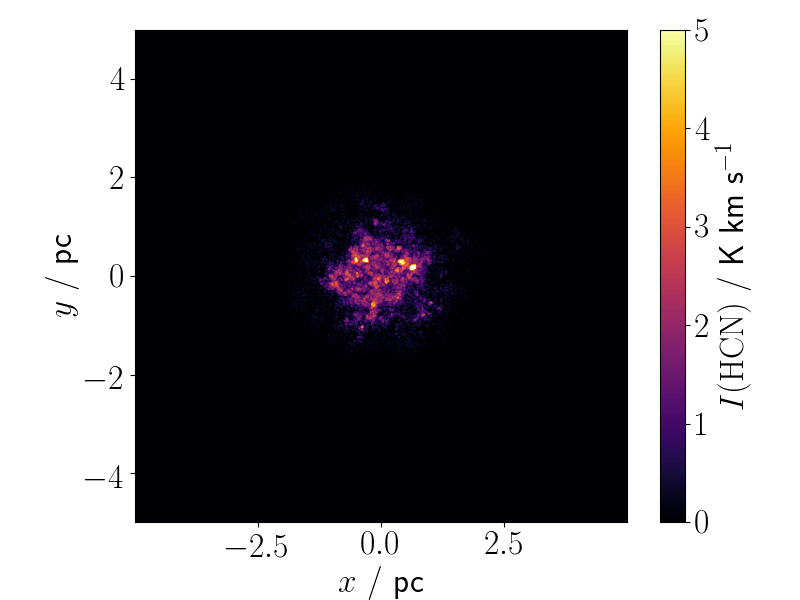}}\quad\\
  \subfigure{\includegraphics[width=0.3\textwidth]{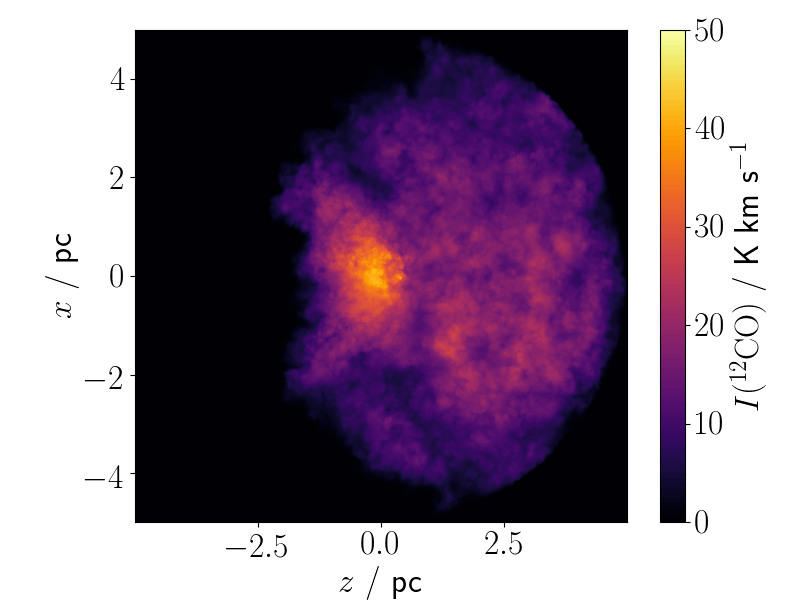}}\quad
  \subfigure{\includegraphics[width=0.3\textwidth]{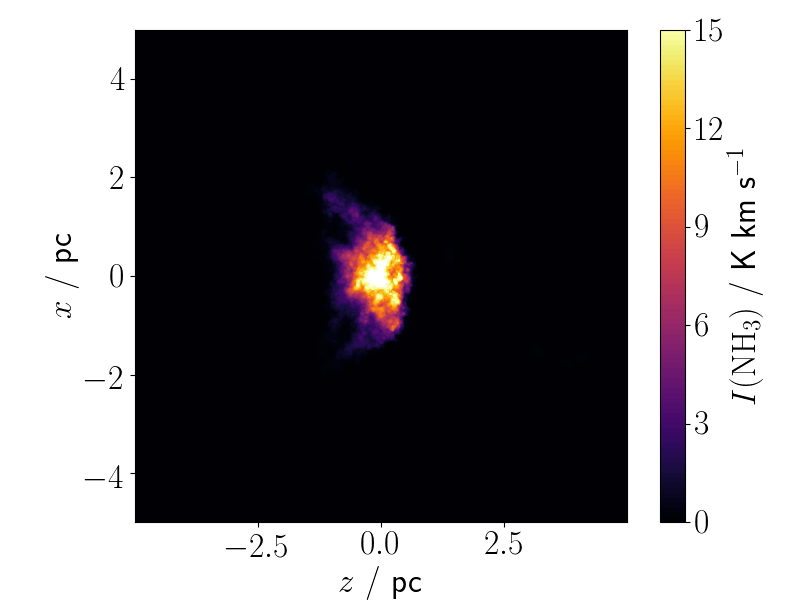}}\quad
  \subfigure{\includegraphics[width=0.3\textwidth]{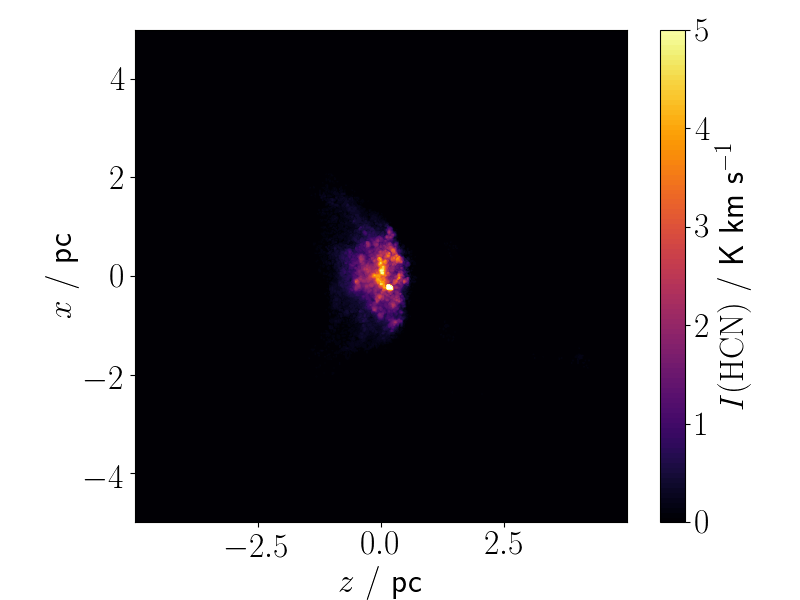}}\quad
  \caption{Integrated intensity maps of {the CO line {\it (left panels)}, the NH$_3$ line {\it (middle panels)}, and the HCN line {\it (right panels)}, from the cloud-collision simulation at $t_{\rm end}=0.49\myr$, looking along the collision axis {\it (top row)} and perpendicular to the collision axis {\it (bottom row)}.}}
  \label{fig:molobs}
\end{figure*}

\begin{figure*}
  \centering
  \subfigure{\includegraphics[width=0.3\textwidth]{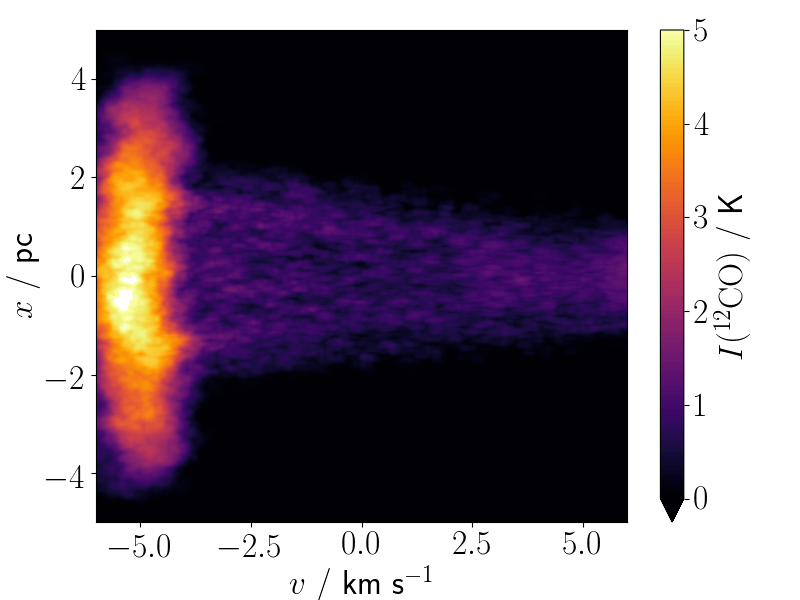}}\quad
  \subfigure{\includegraphics[width=0.3\textwidth]{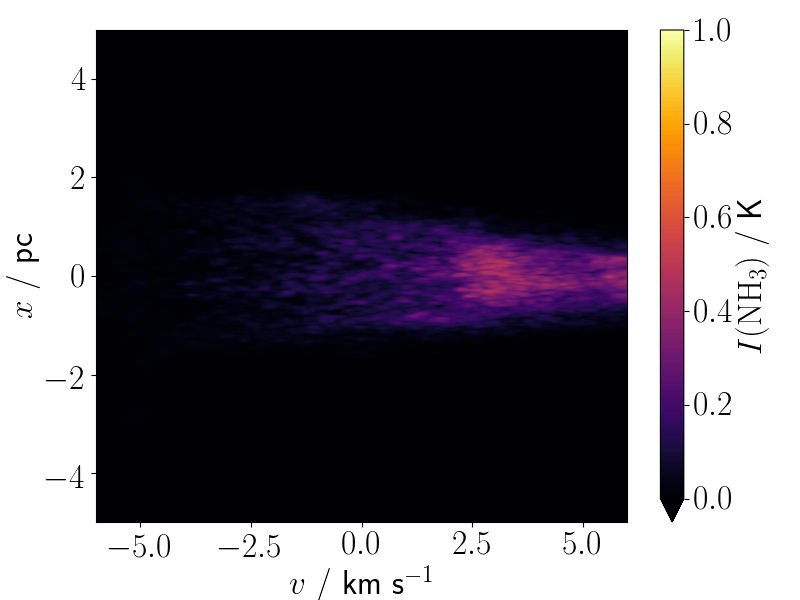}}\quad
  \subfigure{\includegraphics[width=0.3\textwidth]{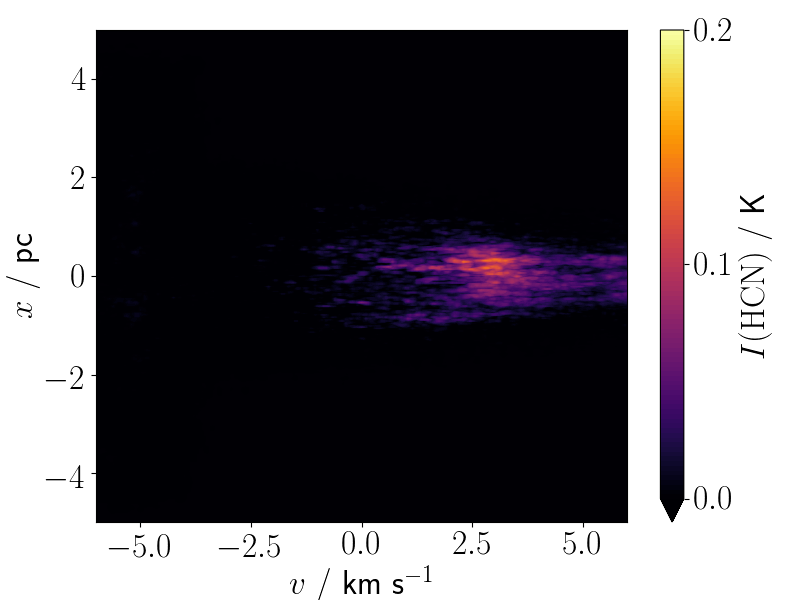}}\quad\\
  \subfigure{\includegraphics[width=0.3\textwidth]{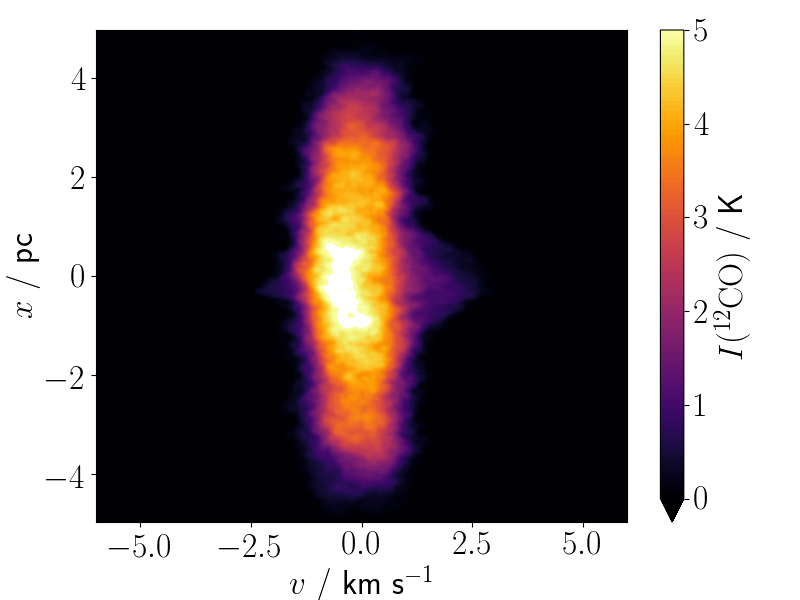}}\quad
  \subfigure{\includegraphics[width=0.3\textwidth]{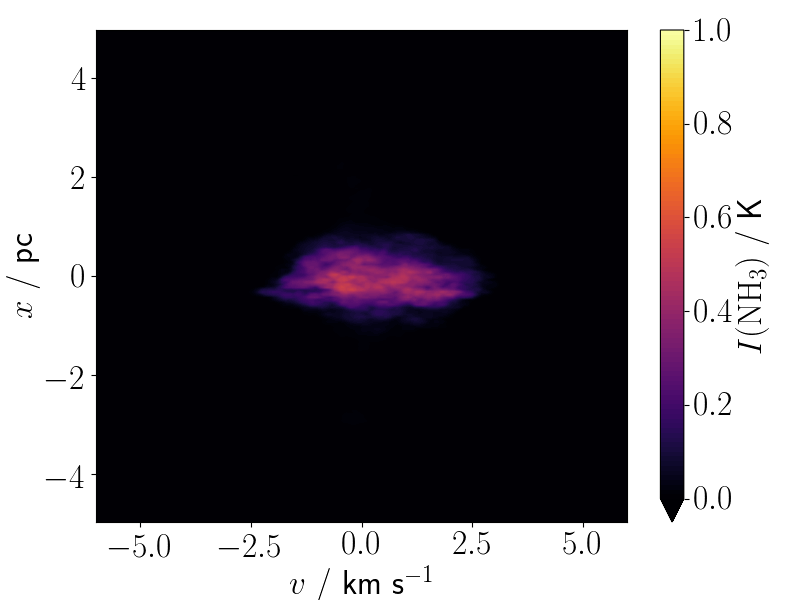}}\quad
  \subfigure{\includegraphics[width=0.3\textwidth]{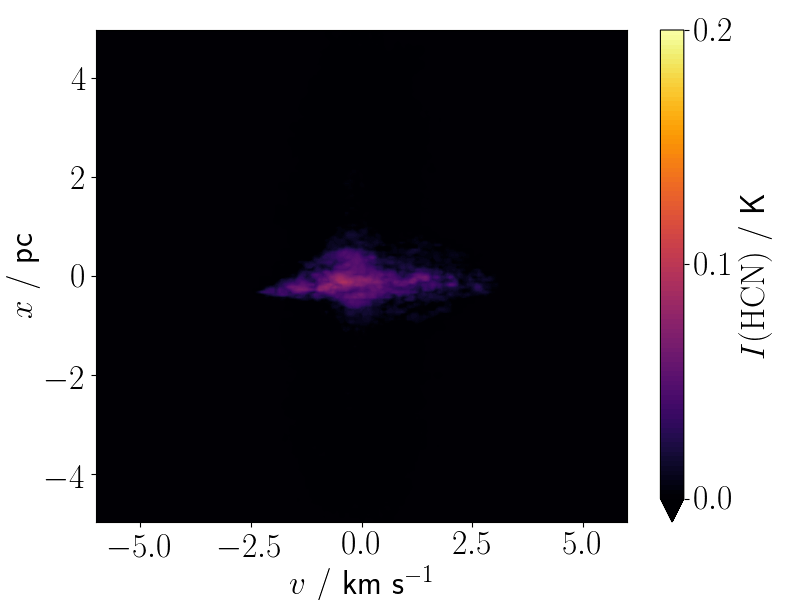}}\quad\\
  \subfigure{\includegraphics[width=0.3\textwidth]{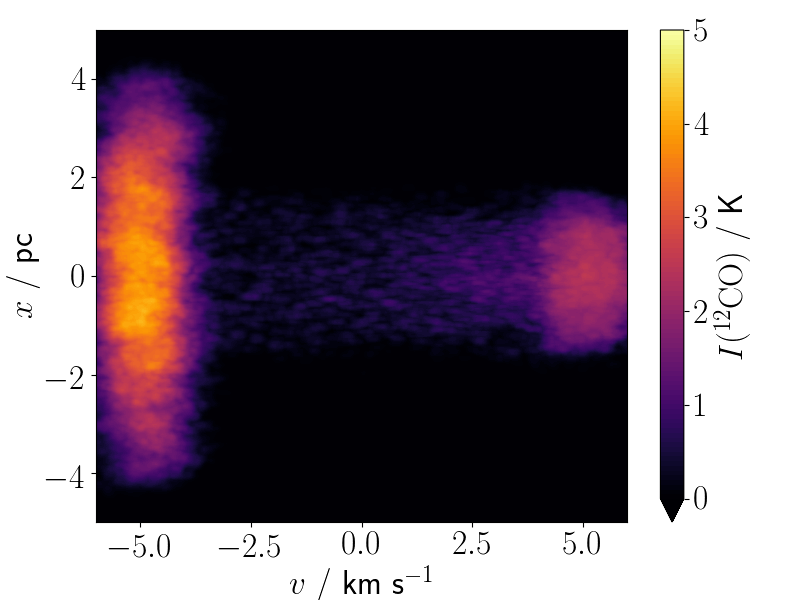}}\quad
  \subfigure{\includegraphics[width=0.3\textwidth]{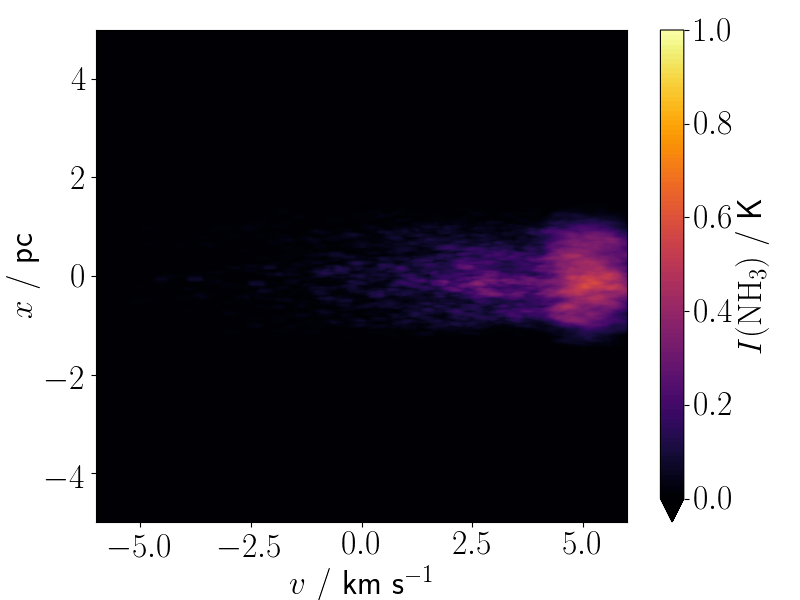}}\quad
  \subfigure{\includegraphics[width=0.3\textwidth]{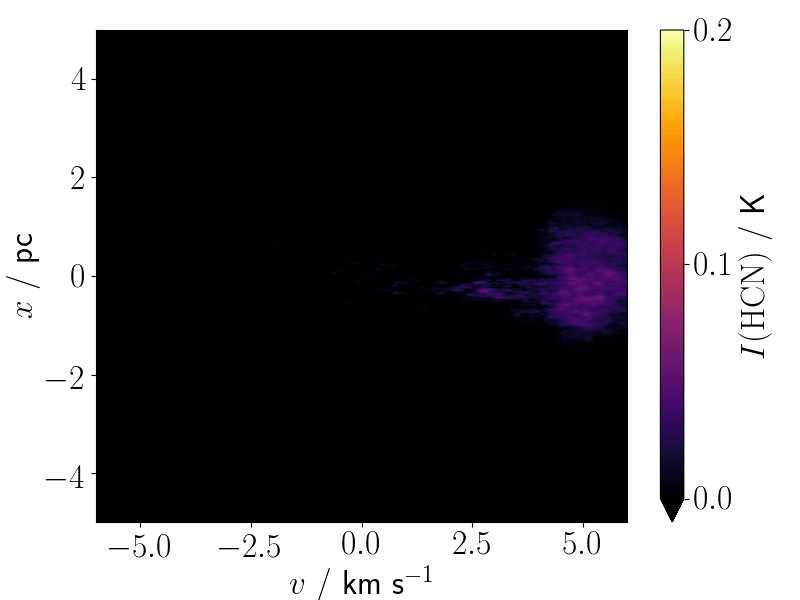}}\quad\\
  \subfigure{\includegraphics[width=0.3\textwidth]{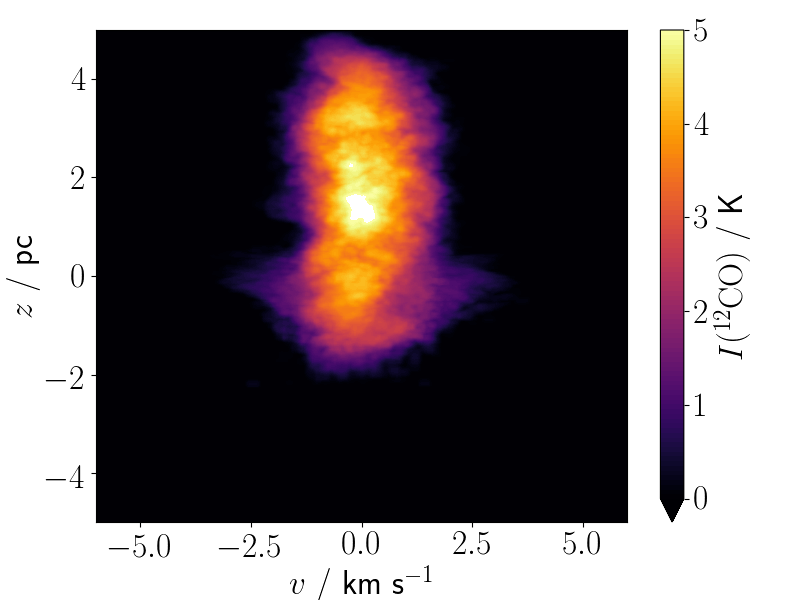}}\quad
  \subfigure{\includegraphics[width=0.3\textwidth]{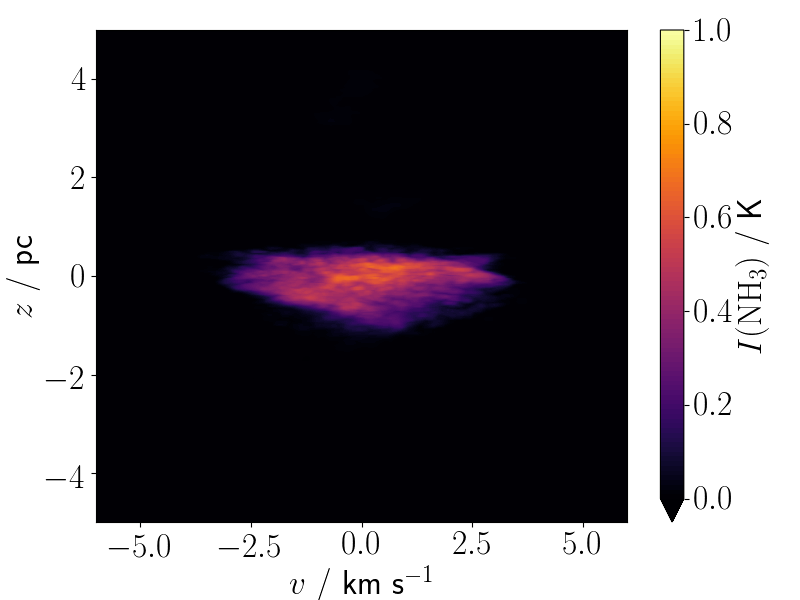}}\quad
  \subfigure{\includegraphics[width=0.3\textwidth]{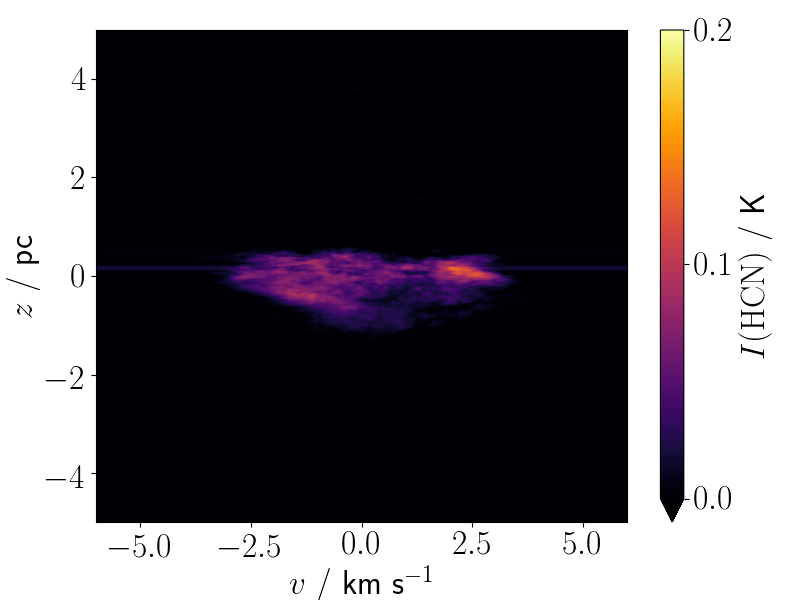}}\quad
  \caption{Position-velocity intensity maps of the CO line {\it (left panels)}, the NH$_3$ line {\it (middle panels)}, and the HCN line {\it (right panels)}. {\it First row:} the cloud-collision simulation at $t_{\rm end}=0.49\myr$, looking along the collision axis. {\it Second row:} the chance-alignment simulation at $t_{\rm end}=0.49\myr$, looking along the collision axis. {\it Third row:} the cloud-collision simulation at $t_{\rm end}=0.24\myr$, looking along the collision axis. {\it Fourth row:} the cloud-collision simulation at $t_{\rm end}=0.49\myr$, looking perpendicular to the collision axis. The intensities have been averaged along the $y$ axis, {to improve the signal-to-noise.}}
  \label{fig:pv}
\end{figure*}

\begin{figure*}
  \centering
  \subfigure{\includegraphics[width=\columnwidth]{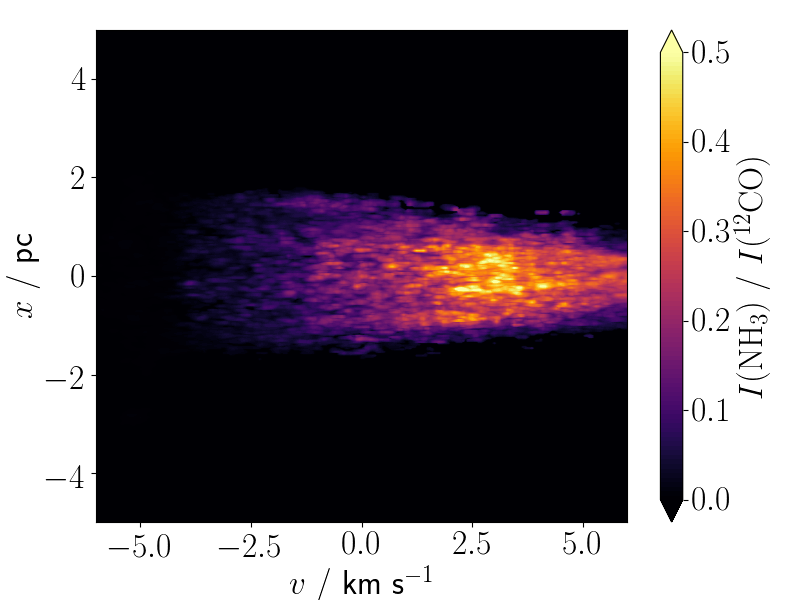}}\quad
  \subfigure{\includegraphics[width=\columnwidth]{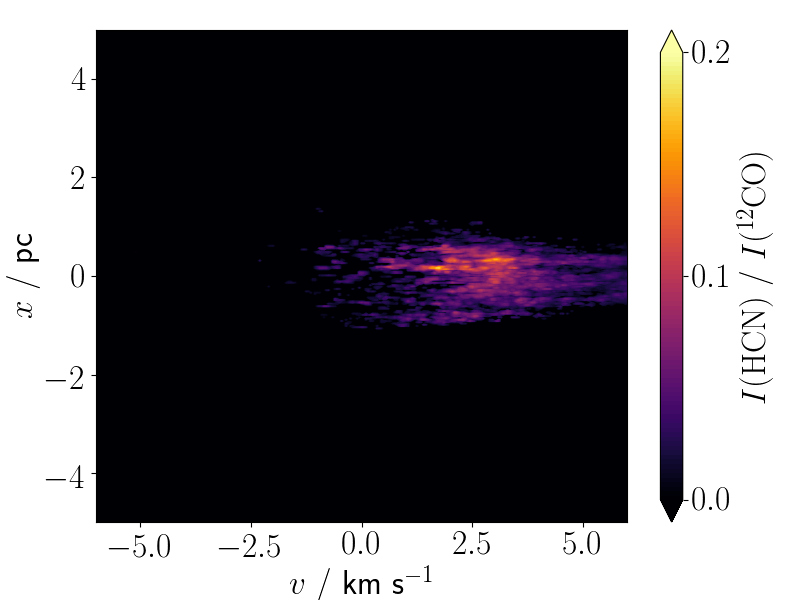}}\\
  \subfigure{\includegraphics[width=\columnwidth]{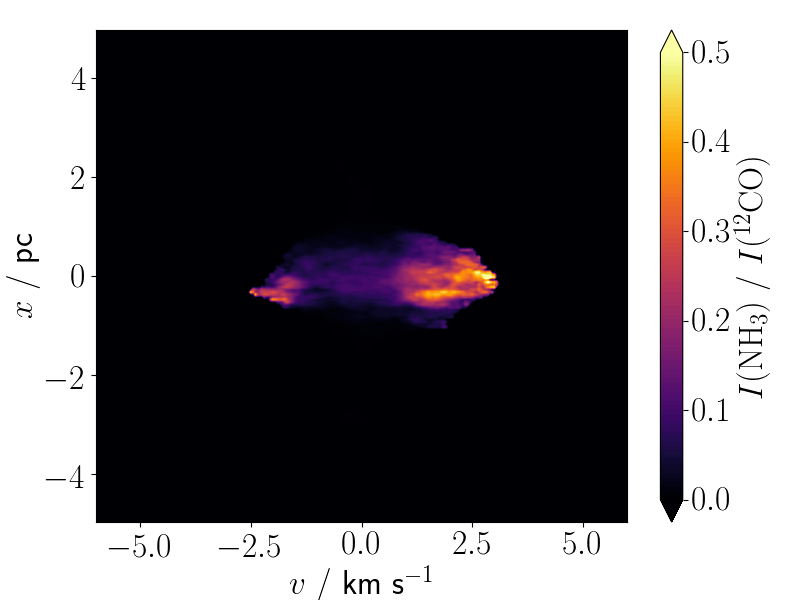}}\quad
  \subfigure{\includegraphics[width=\columnwidth]{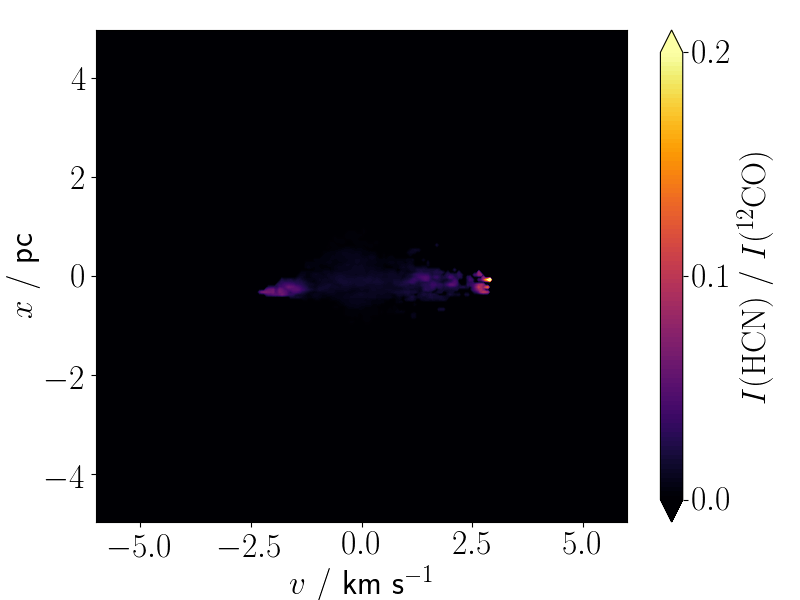}}\\
  \subfigure{\includegraphics[width=\columnwidth]{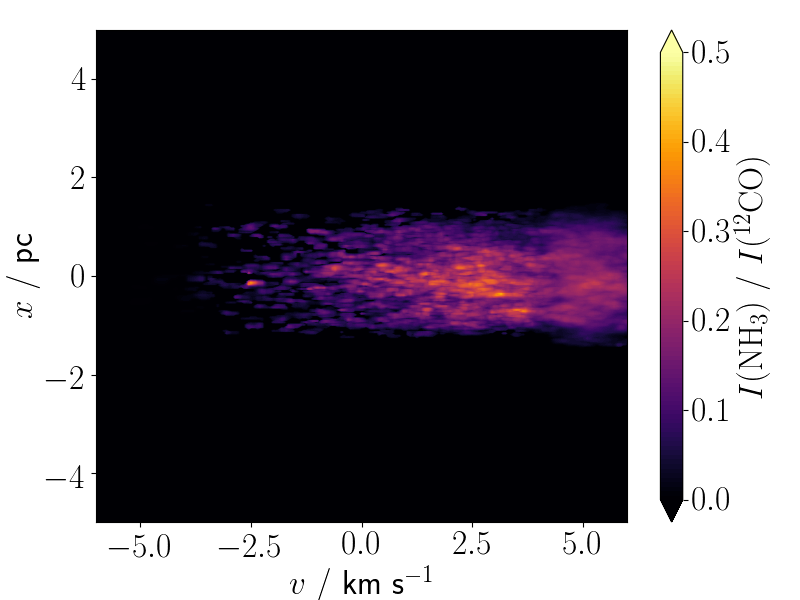}}\quad
  \subfigure{\includegraphics[width=\columnwidth]{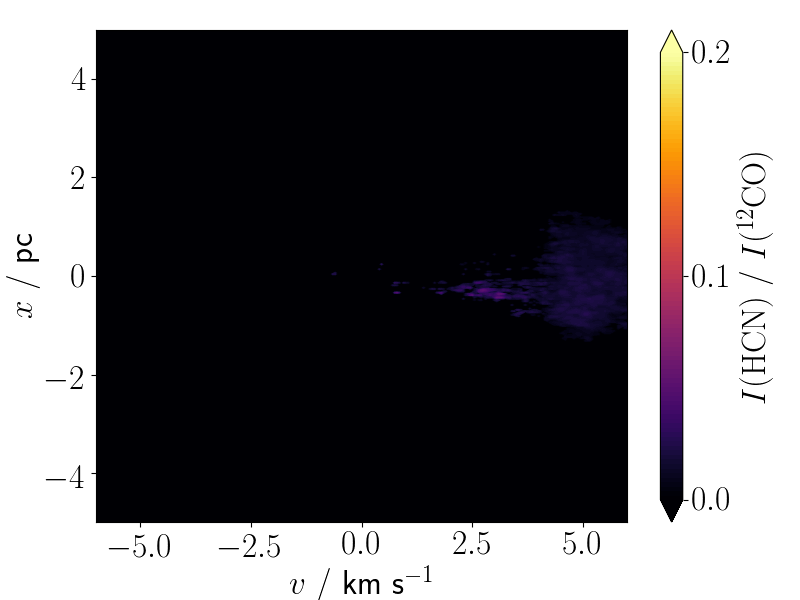}}\\
  \caption{Position-velocity intensity-ratio maps of the NH$_3$ line {\it (left panels)} and the HCN line {\it (right panels)}, both relative to the CO line. {\it First row:} the cloud-collision simulation at $t_{\rm end}=0.49\myr$, looking along the collision axis. {\it Second row:} the chance-alignment simulation at $t_{\rm end}=0.49\myr$, looking along the collision axis. {\it Third row:} the cloud-collision simulation at $t_{\rm end}=0.24\myr$, looking along the collision axis. The intensities have been averaged along the $y$ axis, {to improve the signal-to-noise.}}
  \label{fig:ratio}
\end{figure*}

\section{Method}\label{SEC:Method}

{We perform hydrodynamical simulations using {\sc phantom} \citep{price2018}, a smoothed-particle hydrodynamics (SPH) code.} We consider {a head-on collision} between two {uniform-density} spherical clouds: {a large cloud with radius $R_{\rm large}=5\pc$, mass $M_{\rm large}=10^4\msun$, and density $\rho_{\rm large}=1.3 \times 10^{-21}\gcc$; and a small cloud with $R_{\rm small}=2.5 \pc$, $M_{\rm small}=2.5 \times 10^3 \msun$, and $\rho_{\rm small}=2.6 \times 10^{-21} \gcc$. Assuming a fractional hydrogen abundance by mass of $X=0.70$, the corresponding molecular hydrogen densities are $n_{\rm large}=280\,{\rm H_2\,cm^{-3}}$ and $n_{\rm small}=560\,{\rm H_2\,cm^{-3}}$. These} {cloud masses are somewhat smaller than the representative $\sim 10^5 \msun$ found by \citet{dobbs2015} {in galactic-scale simulations}, but more in line with those in previous theoretical studies of {individual cloud collisions \citep[e.g.][]{haworth2015} and their role in triggering star formation \citep[e.g.][]{balfour2015}.}} We model the initial clouds with randomly positioned SPH particles, each having mass $m_{_{\rm SPH}}\!=\!0.05\,{\rm M}_\odot$. This results in a non-uniform density distribution, with {fractional fluctuations $\Delta\!\ln(\rho)\sim 0.7$} about the mean. The gas in the clouds is isothermal, with sound speed $c_{_{\rm S}} = 0.2 \kms$, corresponding to a gas-kinetic temperature of $10 \kel$ if the hydrogen is molecular. {In addition,} we introduce a turbulent velocity field with power spectrum $P(k) \propto k^{-4}$ and root-mean-squared velocity $2.4 \kms$ (i.e. Mach 12), {consistent with the observed turbulent properties of molecular clouds \citep{larson1981}}.

{At the outset, the cloud centres are at $z_{\rm large}=+R_{\rm large}$ and $z_{\rm small}=-R_{\rm small}$, so that the two clouds are initially just touching.} The rest of the computational domain (a cubic periodic box with side length $80 \pc$) is filled with an ambient medium having density 100 times lower, and sound speed 10 times higher, than in the cloud; this ensures approximate pressure balance across the boundary of the cloud. {In the `cloud-collision simulation',} the clouds {have} {initial velocities of {$v_{\rm large}=-\,5 \kms$ and $v_{\rm small}=+\,5 \kms$} along the $z$-axis, giving a collision velocity of {$\vcol =10\kms$,} comparable to the more violent (and presumably more readily observable) collisions {reported by} \citet{dobbs2015}, and again in line with previous studies of observational {cloud-collision} signatures \citep{haworth2015,bisbas2017}. {As a consequence, a dense, shock compressed layer forms between the clouds. This material has a range of velocities between $\sim v_{\rm large}$ and $\sim v_{\rm small}$, but is concentrated in the range $1.5\kms \la v\la 3.5\kms$, which is due to conservation of momentum and the fact that the material in the small cloud is twice as dense as that in the large cloud.

For the purpose of comparison, we also perform a `chance-alignment simulation', in which $v_{\rm large}=v_{\rm small}=\vcol =0\kms$,} i.e. the clouds evolve independently {(that is, apart from their mutual gravitational attraction). {We follow the evolution of both the cloud-collision simulation, and the chance-alignment simulation,} for a time $t_{\rm end}=R_{\rm small}/v_{\rm small} = 0.49 \myr$.} We introduce sink particles {above a threshold density $\rho_{\rm sink}=10^{-14} \gcc$, with an accretion radius $r_{\rm sink}=0.01 \pc$,} following the prescription in \citet{price2018}.}

{In the cloud-collision simulation,} we track the chemical evolution of all {particles within $5 \pc$ of the origin at {$t_{\rm end}$ (and for the chance-alignment simulation, within $10 \pc$ of the origin at $t_{\rm end}$)}} using {\sc uclchem} \citep{holdship2017}, a time-dependent gas-grain code, with the UMIST12 reaction network \citep{mcelroy2013} and the high-metal elemental abundances from \citet{lee1998}. We assume the standard interstellar background radiation field of \citet{habing1968}. To account for attenuation by dust, we {set} the average column density shielding a gas particle {to
\begin{equation}\label{eqn:NH}
  N_{_{\rm H_2}} = n_{_{\rm H_2}}c_{_{\rm S}}\left(G \rho\right)^{-1/2},
\end{equation}
}i.e. the local density multiplied by the Jeans length, following \citet{priestley2020}. We {then convert $N_{_{\rm H_2}}$} to a visual extinction using {$\av = N_{_{\rm H_2}}/(3.3\times 10^{21}{\rm cm^{-2}})$} \citep{bohlin1978}.

We use the results of our chemical modelling to {calculate molecular line intensities using} {\sc lime} \citep{brinch2010}, a line radiative transfer code. Dust optical properties are taken from \citet{ossenkopf1994}, and molecular data from the {\sc lamda} database \citep{schoier2005}. We use $50\,000$ uniformly distributed sample points, and assign to each sample point the properties of the nearest SPH particle. Increasing the number of sample points incurs a prohibitive computational overhead, and does not substantially change our results. Our output position-position-velocity cubes have a spatial resolution of $0.025 \pc$ per pixel, and cover velocities from $-6$ to $6 \kms$ in $0.05 \kms$ bins.

\section{Results}

{Figure \ref{fig:dens} shows the surface density of the colliding clouds {at $t_{\rm end}=0.49\myr$,} viewed along and perpendicular to the collision axis. {From Figure \ref{fig:dens}, we see (a) that the shock-compressed layer has a lateral extent $\Delta x\sim\Delta y\sim 4\pc$, and (b) that the shocked gas will have a range of velocities, firstly because the unshocked gas at different distances from the collision axis hits the shock at different angles, and secondly because the shocked gas expands away behind the shock. Moreover,} regardless of viewing angle, the gas density is dominated by the {dense material in the shock-compressed layer} between the colliding clouds. This results in significant differences in the appearance of the clouds {when mapped in different} molecular lines, depending on the abundance of the molecule in question and the excitation conditions of the transition. 

{For example,} Figure \ref{fig:molobs} shows the {corresponding} integrated intensity maps for the $^{12}$CO($J\!\!=$1-0), NH$_3 (1,1)$, and HCN($J\!\!=$1-0) {lines,\footnote{In the following we will refer to these lines simply as the CO, NH$_3$ and HCN lines.} at the same time ($t_{\rm end}=0.49 \myr$), looking} along {and perpendicular to} the collision axis. Whilst there is significant $^{12}$CO emission from the {cloud} material that has not yet been {shocked,} the NH$_3$ and HCN lines preferentially trace denser gas {in the shock-compressed layer.} We note that no molecular line directly traces the underlying gas density (i.e. with integrated intensity linearly {proportional to} surface density).}

{Figure \ref{fig:pv} shows PV {maps of the same} three molecular lines {(CO, NH$_3$, HCN)}. The line intensities {have been} averaged along the $y$ axis to compensate for the limited resolution of our {\sc lime} models, {and to improve the signal-to-noise}. {Viewed along the collision axis,} the CO emission shows the bridging feature identified by \citet{haworth2015} {for the cloud-collision simulation}; both the large and small clouds are visible as peaks in the line intensity, separated in velocity space, with {weaker} emission at velocities {between} the two peaks, {due to the gas in the shock-compressed layer}. Molecules tracing denser gas than CO have very different PV morphologies. The large cloud is invisible in both NH$_3$ and HCN emission, and even the small cloud {(with its higher density) is barely visible in HCN}. Only the dense gas {in the shock-compressed layer} shows substantial emission {from NH$_3$ and HCN.}}

{Figure \ref{fig:pv} also shows} PV maps of the CO, NH$_3$ and HCN lines for the chance-alignment simulation, {seen} at the same time ($t_{\rm end}=0.49 \myr$) {and along the same axis as the cloud-collision simulation}, so that the two clouds are aligned one behind the other. As in the cloud-collision simulation, the NH$_3$ and HCN intensity peaks still highlight the denser regions of the clouds, preferentially located towards the cloud centres. However, unlike the cloud-collision simulation, the ranges of velocity for the NH$_3$ and HCN lines are essentially the same as for the CO line. This is because the NH$_3$ and HCN emission is due to denser than average turbulent elements in the clouds that have essentially the same velocity dispersion as the more widely distributed CO emitting gas. We therefore suggest that this comparative analysis of line {intensities} from molecules with different characteristic densities \citep{shirley2015} {constitutes} an additional observational signature of {an ongoing} cloud collision.

{This is even more clearly visible in PV maps of the line intensity ratios, as shown in Figure \ref{fig:ratio}. {Whilst the ratio of NH$_3$ emission to CO emission is $<\!0.1$ in the large cloud and $\sim 0.3$ in the denser small cloud {in the cloud-collision simulation}, it reaches values $\sim 0.5$ at the intermediate velocities of the shock-compressed layer. Similarly, the ratio of HCN emission to CO emission is $<\!0.02$ in the large cloud and $\sim 0.06$ in the small cloud, but reaches values $\sim 0.15$ at the intermediate velocities of the shock-compressed layer.} {While the chance-aligment simulation shows comparable maximum values of the intensity ratios for both molecules, these are not localised in a particular region of PV space, but occur over the same range as the CO emission.}

{PV maps of the line intensities and intensity ratios} for the colliding-cloud simulation at the earlier time of $t_{\rm end}/2\!\sim\!0.24 \myr$ {are shown in Figures \ref{fig:pv} and \ref{fig:ratio} respectively}. At this point, the CO bridging feature is relatively indistinct, but the enhancement in emission from NH$_3$ and HCN is still apparent. The strength of the signature, both in absolute terms and relative to the CO intensity, is reduced, but is still clearly distinct from the {chance-alignment simulation}, where intensity peaks of CO and denser gas tracers are {closely} correlated.}

{Some elements of the signatures explored above still obtain,} even when the cloud-collision simulation is viewed perpendicular to the collision axis. {{The integrated line intensities from the cloud-collision simulation at $t_{\rm end}\!=\!0.49\myr$, viewed perpendicular to the collision axis, are shown in Figure \ref{fig:molobs}}. The compressed layer is highlighted in all three molecules,} but particularly in those tracing denser gas {(NH$_3$ and HCN). It may therefore sometimes be possible to trace the bow shock where a denser smaller cloud is ploughing into a more rarefied larger cloud. However, in this situation there is essentially no velocity information to support the identification of a cloud collision.}

{Figure \ref{fig:pv} shows PV maps of the CO, NH$_3$ and HCN lines from the cloud-collision simulation at $t_{\rm end}\!=\!0.49\myr$, where the position (P) axis is the collision axis, i.e. the $z$ axis. In this case the observer is looking along the $y$ axis, and the intensities have been averaged along the $x$ axis to improve the signal to noise. The velocity range for the NH$_3$ and HCN emission from the cloud-collision simulation viewed perpendicular to the collision axis is a little larger than for the chance-alignment simulation, {but this systematic difference only obtains here because the individual clouds involved are identical apart from their initial bulk velocities.} The $z$-range of the NH$_3$ and HCN emission relative to the $z$-range of the CO emission cannot be used to distinguish the cloud-collision simulation from the chance-alignment simulation: for the cloud-collision simulation the relative $z$-ranges depend on the relative masses and densities of the two clouds involved in the collision; for the chance-alignment simulation the relative $z$-ranges depend on the positions of the two clouds (which are not necessarily exactly aligned relative to the observer, as we have assumed here).}

\section{Discussion \& Conclusions}\label{SEC:Discussion}

{{Since} the enhanced {dense-gas} emission from NH$_3$ and HCN occurs in the same region of PV space as the CO bridging feature, identifying this {enhanced dense-gas emission} requires that a bridging feature is visible in the first place. {Enhanced dense-gas emission is therefore limited --- as a cloud-collision signature ---}  by the same issues regarding the lifetimes of bridging features, and {their} viewing angle dependency, as discussed in \citet{haworth2015b} and \citet{bisbas2017}. {However, whilst} the atomic fine-structure lines investigated in \citet{bisbas2017} trace essentially the same material as the CO line, molecular lines {like NH$_3$ and HCN, which require higher excitation densities,} are necessarily {strongly} enhanced in the compressed layer {resulting from} a cloud collision. NH$_3$ data is available for many Galactic molecular clouds \citep[e.g.][]{friesen2017}, and lines from other dense gas tracers can be observed simultaneously with CO, using submillimetre telescopes. {Our proposed collision signature should thus be relatively simple to investigate in observational studies.}}

{To summarise,} we have performed combined hydrodynamical, chemical, and radiative-transfer simulations of both {colliding clouds and isolated but aligned} clouds, in order to evaluate molecular line signatures of the underlying dynamics. {We find that molecules tracing dense gas, such as NH$_3$ and HCN, have strongly enhanced emission in the {shock-}compressed layer between colliding clouds, {and therefore serve to highlight} CO `bridging features'. This can be used to confirm that observed bridging features are due to {colliding} clouds, rather than unrelated clouds along the line of sight.}

\section*{Acknowledgements}
{We thank the referee for a constructive report which significantly improved this paper.} We are grateful to Giuliana Cosentino and Tom Haworth for their considerate responses to persistent questioning. FDP and APW acknowledge the support of a Consolidated Grant (ST/K00926/1) from the UK Science and Technology Facilities Council (STFC).

\section*{Data Availability}
The data underlying this article will be shared on request. All software used is publically available.

\bibliographystyle{mnras}
\bibliography{cloudcol}


\bsp	
\label{lastpage}
\end{document}